\def\,{\thinspace}
\def\4151{NGC\,4151}
\def\IUE{{\em IUE\/}}

\def\n5{N\,{\sc v}}

\def\mm{\phantom{$-$}}
\def\mone{\phantom{1}}

\documentstyle[11pt,aaspp4]{article}

\lefthead{B.M.\ Peterson et al.}
\righthead{Cross-Correlation Uncertainties}

\begin{document}

\title{On Uncertainties in Cross-Correlation Lags and the
Reality of Wavelength-Dependent Continuum Lags in Active Galactic Nuclei}

\author{
Bradley M.\ Peterson,\footnote{Department of Astronomy, 
The Ohio State University, 174 West 18th Avenue, Columbus, OH 43210-1106.
Electronic mail: {\sf peterson@astronomy.ohio-state.edu}}
Ignaz Wanders,\footnote{ 
School of Physics and Astronomy, University of St.~Andrews, North Haugh, 
St.~Andrews KY16\,9SS, Scotland, United Kingdom. 
Electronic mail: {\sf iw2,kdh1,sjc2@st-andrews.ac.uk}},
Keith Horne,\footnotemark[2]
Stefan Collier,\footnotemark[2] \\
Tal Alexander,\footnote{Institute for Advanced Study, Olden Lane,
Princeton, NJ 08540. 
Electronic mail: {\sf tal@ias.edu}}
Shai Kaspi,\footnote{School of Physics and Astronomy and The Wise Observatory,
Tel Aviv University, Tel Aviv 69978, Israel. 
Electronic mail: {\sf shai,dani@wise.tau.il}}
and Dan Maoz\footnotemark[4]
}

\begin{abstract}
We describe a model-independent method of assessing the uncertainties
in cross-correlation lags determined from
AGN light curves, and use this method to
investigate the reality of lags between UV and optical continuum
variations in well-studied AGNs. Our results confirm
the existence of such lags in NGC 7469. We find that the continuum
variations at 1825\,\AA, 4845\,\AA, and 6962\,\AA\ follow those
at 1315\,\AA\ by $0.22^{+0.12}_{-0.13}$ days, 
$1.25^{+0.48}_{-0.35}$ days, and
$1.84^{+0.93}_{-0.94}$ days, respectively, based on the centroids
of the cross-correlation functions; the error intervals quoted
correspond to 68\% confidence levels, and each of these
lags is greater than zero at no less than 97\% confidence.
We do not find
statistically significant interband continuum lags in NGC 5548, NGC 3783,
or Fairall 9. Wavelength-dependent continuum lags may be
marginally detected in the case of NGC 4151. However, on the basis
of theoretical considerations, 
wavelength-dependent continuum lags in sources other than NGC 7469
are not expected to have been detectable in previous experiments.
We also confirm the existence of a statistically significant lag between
X-ray and UV continuum variations in the blazar
PKS\,2155$-$304.

\end{abstract}


\setcounter{footnote}{0}
\section{Variability in Active Galactic Nuclei}

Over the last ten years, a number of intensive monitoring
experiments on active galactic nuclei (AGNs) have been
carried out (for reviews, see Netzer \& Peterson 1997 and Peterson 1993).
While the primary purpose of many spectroscopic monitoring
campaigns has been to determine the response of the
broad emission lines to continuum variations and thus
determine the structure and kinematics of the line-emitting
gas through the process of ``reverberation mapping''
(Blandford \& McKee 1982), these campaigns have also provided
an opportunity to search for time delays between
different continuum bands. Such time delays are expected to exist if 
the continuum flux in one waveband is reprocessed continuum 
emission from another waveband, either directly through
irradiation or indirectly through viscous processes of the
emitting plasma. The standard accretion-disk model for AGNs
may exhibit both.

Recently, Wanders et al.\ (1997) found evidence for such time
delays in the UV spectra of the Seyfert 1 galaxy NGC 7469
obtained with the {\it International Ultraviolet Explorer (IUE)}\/
spacecraft.
It was found that the continuum flux variations around 1700--1800\,\AA\
lag behind those at 1315\,\AA\ by $0.3\pm0.07$\,d.
This result was subsequently supported by 
contemporaneous optical spectra (Collier et al.\ 1998)
that showed that 
the continuum variations around 4825\,\AA\ lag 
behind the 1315\,\AA\ variations by $1.2\pm0.3$\,d, 
and those around 6925\,\AA\ lag by $1.7\pm0.7$\,d.
The increase in time lag with wavelength in NGC\,7469 is strong
evidence for continuum reprocessing models of AGNs in which the
longer-wavelength photons are reprocessed shorter-wavelength photons
originating closer to the central source. 

A legitimate question to ask at this point is why 
such an effect has not been previously reported in the literature,
even though other AGNs have been monitored in a similar fashion?
This question is of particular interest in the case of
the well-studied galaxy NGC 4151, which was monitored 
at an even higher sampling rate than NGC 7469 during a
10-day campaign in 1993 (Crenshaw et al.\ 1996; 
Kaspi et al.\ 1996; 
Warwick et al.\ 1996; 
Edelson et al.\ 1996). Other Seyfert galaxies that have
been well-monitored simultaneously in the UV and optical 
are NGC 5548 (Clavel et al.\ 1991; Peterson et al.\ 1991, 1992;
Korista et al.\ 1995),
NGC\,3783 (Reichert et al.\ 1994; Stirpe et al.\ 1994),
and Fairall\,9 (Rodr\'{\i}guez-Pascual et al.\ 1997;
Santos-Lle\'{o} et al.\ 1997).
None of these studies found a significant lag between different
continuum waveband flux variations. Why were such lags
found for NGC\,7469, but not for the others?

The answer to this question is not trivial.  The detection of
such lags is dependent on the sampling characteristics
of the light curves, as well as on the variations of the light
curves themselves, i.e., the auto-correlation function (ACF) of the
light curve. In any event, the formal
statistical significance of a time-lag detection through
a cross-correlation analysis is hard to assess. There is
no generally agreed upon way to estimate the errors in
cross-correlation lags and thus attach a level of significance
to a time-lag detection. Indeed, interband continuum lags
{\it have} been reported in previous campaigns, but
the detections were not thought to be statistically significant.

In this paper, we will introduce what we believe is a
conservative model-independent approach to estimating the uncertainties in
cross-correlation lag determinations (\S\,2) 
and use it to re-examine existing data sets (\S\,3). 
We will then discuss the implications of these results (\S\,4) and 
present our conclusions (\S\,5).

\section{Estimation of Cross-Correlation Uncertainties}

In studies of AGN time series, we are faced with the
following general situation: we have
two light curves with fluxes $\{A_1, A_2, \ldots, A_N\}$ and
$\{B_1, B_2, \ldots, B_M\}$, that are often irregularly sampled at
times $\{t_1, t_2, \ldots, t_N\}$ and $\{t_1', t_2', \ldots, 
t_M'\}$, respectively. The two light curves have $N$ and $M$
observed fluxes $F_i$ with corresponding estimates of the
1$\sigma$ measurement errors 
$\{\alpha_1, \alpha_2, \ldots, \alpha_N\}$ and $\{\beta_1, \beta_2, \ldots, 
\beta_M\}$. We then cross-correlate the two light curves. The
quantity we seek is the lag (whether it be the lag of the peak
or the centroid of the CCF is irrelevant at this point),
and its uncertainty. In general, we
recommend using the centroid of the cross-correlation
function $\tau_{cent}$ rather than the location of the
peak $\tau_{peak}$. Koratkar \& Gaskell (1991) and 
Penston (1991) have shown that $\tau_{cent}$
is a measure of the luminosity-weighted radius of the reprocessing region.
However, as Robinson \& P\'{e}rez (1990) have shown,
$\tau_{peak}$ cannot be interpreted as a physical
quantity in any simple manner. Besides being model-dependent,
it also depends on the auto-correlation function of the light curves,
as well as the sampling of the light curves.
The results of their simulations showed that the peak lag
was most sensitive to the inner regions of a reprocessing
region (also as noted by Gaskell \& Sparke 1986), 
and thus likely underestimates the actual size of the reprocessing region. 

\subsection{Model-Dependent Monte Carlo Methods}

While there is no general agreement on how to
estimate the uncertainties in cross-correlation lags,
probably the method that has the widest acceptance is
Monte Carlo (MC) simulations. In studies of AGN variability,
these are generally carried out in the following fashion:
first, one adopts a model light curve that is supposed to
``drive'' the other variations. 
Selection of a driving light curve is usually
based on the shortest UV continuum wavelength observed, and
it is assumed that the ionizing continuum (which is supposed to drive the
emission-line variations) behaves in a similar fashion.
The driving light curve is then convolved with a ``transfer
function'' that produces a ``responding'' light curve,
which most often represents the response of a broad emission line
to the continuum variations.
The driving and responding light curves are then 
``sampled'' in a fashion that somehow mimics the
real observations, and then the effects of observational
uncertainties, both random and systematic, are included. These
artificial data sets are then cross-correlated as if they
were real data, and the cross-correlation lag (based either
on the peak or centroid of the cross-correlation function)
is recorded. This process is repeated a large number of
times to build up a cross-correlation peak distribution
(CCPD; see Maoz \& Netzer 1989) which we emphasize
is not necessarily a normal distribution. From the CCPD, the likelihood that
a given lag falls in some particular range can be determined.

The principal problem with this particular method is that the
results are highly model-dependent --- the uncertainty we derive
is only valid to the extent that our models of the driving
light curve and the transfer function are reasonable
approximations to the true situation. 

The behavior of the
continuum is one of the principal sources of uncertainty
in MC simulations. AGNs undergo irregular continuum
variations whose origin is not known, but thought by some to have some
relationship to accretion-disk instabilities or changes in the accretion rate
(e.g., Mineshige \& Shields 1990;
Siemiginowska, Czerny, \& Kostyunin 1996).
AGN-like continuum variations can be parameterized by
a power-density spectrum (PDS) of power-law form
$P(f) \propto f^{\alpha}$ (where $f$ is the temporal frequency in Hz
or days$^{-1}$)
and index $\alpha \approx -2\pm0.5$.
A model continuum that bears a close resemblance to observed continua
can therefore be generated from a PDS of the correct form by
randomizing the phases of the Fourier components, performing
the inverse Fourier transform, and then normalizing the continuum
to yield the desired fractional variability ($F_{var}$,
as defined by Rodr\'{\i}guez-Pascual et al.\ 1997). 
For a limited data train, the variations in such model continua
can show a very wide variety of structures. Indeed, for well-sampled
high signal-to-noise ratio light curves, the principal source
of variance among various MC realizations is often
determined by differences in continuum light curve features,
even when all the model continua are characterized by the same
PDS power-law index and fractional variability. In such cases,
MC simulations can in fact {\em overestimate} the
uncertainty in cross-correlation lags. Certainly, however, such simulations
are extremely valuable and valid in experiment design, since the
particulars of the continuum behavior cannot be predicted.

Determination of a model responding light curve is similarly
fraught with uncertainty, not only because of the unknown geometry
of the reprocessing region, but because the reprocessing physics needs to
be modeled as well (regardless of whether we are considering
reprocessing into emission lines or continuum radiation). 

We wish to consider, therefore, methods by which the uncertainties in
cross-correlation lags can be estimated in a less model-dependent
fashion that is simple to implement.

\subsection{Model-Independent Monte Carlo Methods}

Given a sufficiently long, continuously sampled light curve
with error-free observations, it should be possible to determine
a cross-correlation lag to arbitrary precision. The experimentally
determined lag is not necessarily highly repeatable in time
(even if the transfer function does not change with time) simply
because the specific pattern of continuum variability can change
with time; since the responding light curve is a convolution of
the transfer function and the continuum light curve, the line
response can also show pronounced differences in repeat experiments.
It is simple to show (e.g., Penston 1991) that the cross-correlation
function itself is a convolution of the transfer function and
the driving continuum auto-correlation function, so the centroid
or peak of the cross-correlation function can vary on account of
differences in the driving continuum autocorrelation function.
We therefore consider only sources of uncertainty that
are introduced by deficiencies in the experimental data,
and not the broader problem of attempting to determine
limitations on the transfer function imposed by the measurements.

Two of the principal
sources of uncertainty in cross-correlation lags 
are (a) flux uncertainties in individual measurements
and (b) uncertainties associated with the observational  sampling
of the light curves, i.e., the intervals between observations
and the duration of the experiment. We consider  methods for
assessing the uncertainties associated with each of these.

In many AGN monitoring programs, uncertainties in measured fluxes
can be a significant source of error. The importance of
flux errors can be easily assessed through MC simulations.
One just takes each real measurement $A_i$ and alters it
by adding a random noise contribution. We assume that the errors
in fluxes are normally distributed, and we thus modify each
flux by random Gaussian deviates based on the quoted error $\alpha_i$ for
each datum. The modification of each data point is statistically
independent from each of the others. In a single MC
realization, each data point is modified, the CCF is computed,
and the lag is recorded. Multiple realizations build up the
CCPD. Comparing a large number of independent realizations should
reveal that the average value of an individual point
remains $A_i$, and the standard deviation should be $\alpha_i$.
We will refer to this
process as ``flux randomization'' (FR). 
This procedure is commonly used as part of MC simulations, as described above.

Cross-correlation results can be highly sensitive to 
individual data points or even combinations of data points.
We can, then, to some extent test the sensitivity of a cross-correlation
result by considering only subsets of the original parent data set.
This process is similar, but not identical, to a commonly used
and powerful statistical technique known as ``bootstrapping''.
The bootstrap method can be used to evaluate the
significance of correlations based on limited data
(see Press et al.\ 1992 for a brief description). For example,
consider a set of observations $\{(x_1,y_1), (x_2,y_2),
\ldots, (x_N,y_N)\}$. For a single realization, we draw from this sample $N$ 
pairs of randomly selected points $(x_i,y_i)$ 
{\em without regard to whether or not they have been previously
selected}. This results in exclusion of some data pairs, and
counts others multiply. The correlation analysis is again performed,
and the correlation coefficient is recorded. Multiple realizations
lead to a mean and standard deviation for the correlation coefficient.

Extension of this process to time series, in which one set of values
is highly ordered, is not obvious. The temporal order of the data points must
be preserved. But we can still select a subset of the original
data points without regard to previous selection; how to incorporate
these multiple data points into the analysis becomes somewhat
problematic: one can either ignore the redundant data points
(which effectively reduces the size of the selected sample, typically by 
a factor $\sim 1/e\approx0.37$, which is the Poisson probability
of not selecting any particular point), or one can increase the weight of the 
multiply selected data points. The latter can be used in cross-correlation
methods that make use of the flux uncertainties, such as 
the discrete correlation function (DCF) method (Edelson \& Krolik 1988) or
the $z$-transformed discrete correlation function (ZDCF) method
(Alexander 1997). Here we will use only the
interpolated cross-correlation function (ICCF) method, which does not
make use of the flux errors in the data, and therefore we will simply
{\em exclude}\/ redundant selections from the simulations.
Thus, each realization is based on a randomly chosen subset of the original
data points, and we will refer to this as ``random subset selection''
(RSS). We argue that this method gives a fairly conservative estimate
of the errors due to sampling: the effects of individual data points
are accounted for statistically by random removal of such points, and
each MC realization is based on a number of data points that
is typically 37\% smaller than the real data set, so the individual
MC realizations are likely to produce a broader range of
values than similar sets the same size as the original set.

The two elements that we describe above are complementary in that they
test the sensitivity of the cross-correlation results to
flux errors (FR) and to sampling characteristics (RSS), and
the two processes can be combined in a single simulation.
This combination does not account for {\em all} possible sources
of error, however, as we still cannot test what might have occurred
in the gaps between observations. In all realizations, 
the data are restricted to some subset of the
actual observation times $\{t_1, t_2, \ldots, t_N\}$ and
$\{t_1', t_2', \ldots, t_M'\}$. Dependence on the times of
observations can be tested only if one has a model for
the flux behavior at other times --- this leads to 
model-dependent simulations of the sort described above.
Our proposed method should be valid, however, to
the extent that the ICCF method is itself valid, i.e., the
gaps between observations must be sufficiently small that low-order
interpolation between them is a reasonable approximation to the
true  behavior of the light curve.

We also caution the reader that the method as implemented here
does not take into account systematic errors that may be
operative in some cases. An obvious example is the problem
of ``correlated errors'', which can arise when multiple measurements
are made from a single spectrum, which is the case for many of
the data discussed here. Suppose, for example, that the
flux-calibration of a set of spectra has a random-error component.
Then a time series of measurements of a even a constant source will show
(presumably small) variations, with all measured spectral
features varying similarly and in phase. Cross-correlation of these
time series will yield a peak at zero lag. In the case of
AGN variability studies, correlated errors have been found to 
introduce significant systematic effects only in one case,
Akn 120 (Gaskell \& Peterson 1987). This  was a result of the
relatively small equivalent width of the [O\,{\sc iii}]$\lambda5007$
emission line (which is commonly used for flux calibration in
optical variability studies of AGNs) and the relatively large
measurement uncertainties (typically $\sim$8\%) that could be
obtained with pre-CCD era detectors. On the basis of simulations,
Gaskell \& Peterson (1987) concluded that the effects of correlated errors 
would be obviated by a modest improvement in the quality of the
data, and this has indeed been borne out (Peterson et al.\ 1998).
As a cautionary measure, we have examined some of the data sets used
here for correlated errors. We have looked for
correlations between the differences of pairs of measurements
closely spaced in time, i.e., correlations between
$A(t_j) - A(t_i)$ and $B(t_j) - B(t_i)$, where $A$ and $B$ are
measured from the same spectra. We find no evidence that
correlated errors affect the data discussed here. In any event,
we note that if correlated errors were in fact present, the
artificial correlation at zero lag would tend to drive our
measured cross-correlation lags to {\em smaller} values,
thus masking the very effect we are attempting to measure.
If correlated errors affected these data, our results would
thus be conservative estimates of the probability that the
interband lags are greater than zero.

\subsection{Test Results}

In order to test our assertion that the combined FR/RSS method will
yield conservative error estimates of cross-correlation lags,
we will undertake a number of numerical tests based on 
simple models, which we construct as
follows: first a driving continuum is constructed  from a
power spectrum of a given PDS index $\alpha$ and randomized phases.
This process is described
more fully by White \& Peterson (1994). The spectrum produced is then
normalized to a root-mean-square (rms) variability level
$F_{var}$ that is consistent with experimental results. The
driving continuum is then convolved with a specified transfer
function to produce a responding light curve, which can then be
renormalized to a value of $F_{var}$ that is consistent with
what is observed for emission lines. These two light curves 
constitute ``parent'' light curves from which individual experimental
realizations can be drawn. 

For each realization, we select according
to some specified procedure $N$ data points spanning some time
span $t_N - t_1$ from both the driving  and
responding (which we will call respectively the ``continuum'' and ``
emission-line'') light curves. In these tests, the
continuum and emission-line fluxes are always paired, i.e., both
time series have $N$ points sampled at the same time; this
is not a necessary restriction, but it mimics what occurs in nearly
all real data sets. 
Once a sample of data points has been selected, the effects
of random errors are included.
Fractional flux deviations  based on random 
Gaussian deviates with variances again chosen to be consistent with
real data sets are then used to redistribute the sampled fluxes and
thus simulate the effect of random flux errors. 
The $N$ selected data points from each series are then cross correlated
as if they were real data. A cross-correlation realization is deemed to 
have succeeded if the maximum correlation coefficient $r_{max}$ is
large enough to conclude that the correlation between the two
series is significant at a level of confidence greater than 95\%.
The time delay $\tau_{peak}$ at which $r_{max}$ occurs  and the
centroid of the cross-correlation function $\tau_{cent}$, 
here based on all points
with correlation coefficients in excess of $0.8r_{max}$,
are recorded and used to build up distribution functions (CCPDs) for these
parameters. After a large number of realizations (usually
500 to 2000), this distribution can be integrated to determine
the probability that a given realization will yield a result
above or below some specified value. We reiterate
the important point made by Maoz \& Netzer (1989) that these distributions
are almost always non-normal and the standard deviation of the
distribution is not a good characterization of the uncertainty
in the lag. Unless otherwise noted, the 
uncertainties we will quote, computed directly
from the CCPDs, will be $\pm\Delta \tau_{68}$,
defined such that
68.27\% of the realizations yield results between
$\tau_{median} - \Delta\tau_{68}$ and 
$\tau_{median} + \Delta\tau_{68}$, or equivalently,
15.87\% of the realizations give values below 
$\tau_{median} - \Delta\tau_{68}$ and an equal number
above $\tau_{median} + \Delta\tau_{68}$;
these limits correspond to 1$\sigma$
errors for a normal distribution. A sample CCPD for one of the
tests performed below is shown in Fig.\ 1.

In order to compare the results of our method with the results of
model-dependent MC simulations, we select at random {\em one}
of the realizations for a given model, and we independently attempt
to estimate the uncertainties in the lag by FR/RSS,
and compare these to the results of the model-dependent
simulations.
The results of a single sample test are shown
in Table~1. For the test described here (based on one of the
White \& Peterson 1994 models with a $\Delta T = 1$\,day
interpolation grid and where the 
responding emission-line region is modeled as a
thin spherical shell), the continuum PDS has index $\alpha = -2.5$, and
$F_{var} =0.30$. The BLR is characterized by $R = 20$ light days
and asymmetry parameter $A=0$ (i.e., completely isotropic line emission).
The assigned fractional error levels are $0.035$ for the
continuum and $0.037$ for the line. Each sample consists of
$N=40$ data points, chosen at random within a 200-day span.
Column (1) gives the parameter recorded, either the 
centroid $\tau_{cent}$ or peak $\tau_{peak}$ of the 
cross-correlation function. Column (2) gives the median value and
uncertainties (i.e., $\pm\Delta\tau_{68}$) for the model-dependent MC
simulation. 
Column (3) gives the cross-correlation lag ($\tau_{cent}$ or
$\tau_{peak}$) based on the
single randomly selected realization and the FR/RSS-based
uncertainty computed from that realization. For comparison,
columns (4) and (5) give
respectively the uncertainties based on RSS and FS
alone. The results of this randomly chosen test are 
representative in several respects. First, as expected,
the error estimates based on either RSS or FR alone are generally
smaller than the those for the combined FR/RSS. Second,
the expected value of the lag from the model (20 days) is
within the error limits for each of the methods. Third,
the combined FR/RSS uncertainties are somewhat larger than
the uncertainties based on the model-dependent 
MC simulations, i.e., they seem to provide a 
{\em conservative} estimate of the uncertainties. However,
the uncertainties based on either FR alone or RSS alone are
usually smaller than the MC-based uncertainties, and we will therefore
not consider these further.

As we mentioned earlier, specification of the continuum
with only the PDS index $\alpha$ and the fractional variability
$F_{var}$ can produce a wide variety of light curves. In Table 2,
we show the results from 10 nominally identical simulations,
quoting both the model-dependent Monte Carlo results and
the FR/RSS results based in each case on one randomly selected
MD realization (Test 1 is from Table 1). The entries in Table 2
show that $\tau_{cent}$ and $\tau_{peak}$ vary somewhat, depending
on the precise continuum behavior that is sampled. Moreover,
the uncertainties in these parameters can also vary. 
The FR/RSS uncertainty is always greater than or approximately
equal to the model-dependent Monte Carlo uncertainty, and
the true value of 20 days is virtually always within the
quoted uncertainty range.

In Table 3, we show the results for a similar series of tests
on a thick-shell model with good temporal sampling ($N = 50$ observations
randomly spaced on a grid of spacing $\Delta T = 0.5$\,days
during a 100-day interval). The continuum used is 
characterized by a power-law PDS with $\alpha =2$
and $F_{var} = 0.14$.
The specific thick-shell model
chosen is the Wanders et al.\ (1995) for NGC 5548, with
an inner radius of 1 light day, and outer radius of 12 light days,
and an emission-line asymmetry $A=0.6$ (although we do not here
consider the effects of an anisotropic continuum source,
as did Wanders et al.). Both
lines and continuum are assumed to be measured to 2.5\%. The
results are shown in Table 3. These tests show that
lag can vary significantly depending on the details of the
light curve, which are different for each of the 10 tests shown
here. Again it is seen that the FR/RSS errors
are similar to or slightly greater than the model-dependent MC
errors, even though the errors themselves also vary from
test to test.

In Table 4, we show more extensive comparisons of the model-dependent MC-based
and FR/RSS uncertainties. The models shown in Table 4 are a
subset of the thin-shell models that cover most of the parameter
space considered by White \& Peterson (1994).
Column (1) gives the PDS index for the continuum model and
column (2) gives the broad-line region radius (the transfer function
for a thin spherical shell of radius $R$ is a rectangular function
that is non-zero only in the range 0 to $2R/c$, thus with centroid
$R/c$). Column (3) gives the number of observational data points
sampled from a 200-day span of data --- the mean sampling intervals
are thus about 5 days for $N=40$,
10 days for $N=20$, and 20 days for $N=10$. The cases $N=20$ and
$N=10$ constitute rather poor sampling by AGN monitoring standards,
particularly in the case of the flatter PDS ($\alpha=-1.8$) which
has more rapid continuum variations. Columns (4) and (5) give the
median value of the centroid and associated uncertainties for the
complete model-dependent Monte-Carlo simulations (MC) and
for the model-independent (FR/RSS) simulations, respectively.
The corresponding values of $\tau_{peak}$ are given in columns (6)
and (7). These results show that in general the FR/RSS uncertainties
are quite comparable or slightly more conservative than the
model-dependent MC errors. As the sampling rate is decreased
(i.e., as $N$ decreases), the uncertainties in the cross-correlation
results become very large, as can be seen in the Table. In the
case of very poor sampling, the FR/RSS uncertainties are much larger
than the MC errors because the typical FR/RRS realization is working
with about 37\% fewer points from an already poorly sampled light
curve. Thus, in the limit of very poor sampling, the FR/RSS method fails in
a desirable way by giving error estimates that are too large
(in contrast, say, to the Gaskell--Peterson [Gaskell \& Peterson 1987]
analytic formula,
as shown by White \& Peterson 1994).

In summary, we conclude that FR/RSS gives reliable, conservative
uncertainties for a wide variety of cross-correlation simulations.
We now proceed to apply this method to estimate errors for
measured interband continuum lags for a number of AGNs that
have been simultaneously monitored in the UV and optical.

\section{Wavelength-Dependent Lags in Well-Studied Sources}

In Table 5, we show cross-correlation results for NGC 7469,
where the UV data are from Wanders et al.\ (1997) and the
optical data are from Collier et al.\ (1998). In addition to
the cross-correlation lags $\tau_{cent}$ and $\tau_{peak}$,
we show estimates of the uncertainties in each of these
parameters, based on FR/RSS Monte Carlo simulations  
as described in the previous section. We also
show the probability, again based on the FR/RSS simulations,
that the true lag is less than or equal to zero.
As noted earlier, the
FR/RSS uncertainties are conservative compared to model-dependent
Monte-Carlo simulations, and thus the uncertainties reported here
are somewhat larger than quoted in the original work
(for example, for the 1315\,\AA/1825\,\AA\ cross-correlation,
Wanders et al.\ find $\tau_{cent} = 0.22\pm0.07$\,days for the
TOMSIPS-derived data, and we assign uncertainties of
$+0.12/-0.13$\,days). Even using the FR/RSS errors, the
wavelength-dependent continuum lags are statistically significant
(although not $\tau_{peak}$ for the 1315\,\AA/1825\,\AA\ cross-correlation),
though not highly significant.

A number of additional multiwavelength AGN monitoring campaigns have been
carried out over the past years, and these should provide us with
data suitable for searching for similar wavelength-dependent continuum
lags in other AGNs. Several non-blazar
AGNs have been monitored simultaneously in the optical and
ultraviolet\footnote{All of the non-blazar AGNs studied here
were monitored by the International AGN Watch consortium. The
light curves used in these calculations can be obtained through
the International AGN Watch site on the World-Wide Web at URL
{\sf http://www.astronomy.ohio-state.edu/{\scriptsize $\sim$}agnwatch/}.},
but with the exception of NGC 7469, these have
yielded only upper limits on possible lags between continuum
variations in the short-wavelength ultraviolet (usually at
about 1300\,\AA) and the optical (usually at about 5000\,\AA).
However, the upper limits quoted depend on the uncertainties in
the lag determination, and these have not been estimated in
a uniform fashion. Here we employ the FR/RSS methodology to re-evaluate the 
uncertainties in cross-correlation lags for these well-studied AGNs
in a uniform fashion in order to investigate whether or not
the NGC 7469 results may be atypical. We will also include
for completeness results on one well-studied blazar. The AGNs we consider
are the following:
\begin{enumerate}
\item {\bf NGC 5548:} We re-examine the results of two
UV/optical campaigns, the first from 1989 (Clavel et al.\ 1991,
Peterson et al.\ 1991, 1992) and a more intensive campaign
in 1993 (Korista et al.\ 1995). In the  earlier campaign, the 
\IUE\ data covered an eight-month period from 1988 December to
1989 August, yielding measurements at 60 epochs with an
average interval of about 4 days between observations. Concurrent
ground-based spectra yielded optical continuum measurements at
an average spacing of about 2.5 days. The 1993 program was
anchored by {\it Hubble Space Telescope (HST)}\ 
FOS observations which were obtained once
per day for 39 days. The program was preceded by a series of
\IUE\ observations (one every other day) which together yield a light
curve consisting of 53 data points over a 76-day period. The
concurrent ground-based program produced 105 optical continuum
measurements during the same period.
\item {\bf NGC 3783:} This galaxy was monitored with \IUE\ from
1991 December to 1992 July (Reichert et al.\ 1994),
yielding 69 observations with an average spacing of 3.3 days.
Concurrent ground-based observations (Stirpe et al.\ 1994) 
produced an optical light curve that was comprised of 72 observations
during the same period with approximately the same average spacing
as the \IUE\ data.
\item {\bf Fairall 9:} \IUE\ observations (Rodr\'{\i}guez-Pascual et al.\ 1997)
were obtained between 1994 April and 1994 December. 
Optical observations were obtained from the beginning of this program 
through the end of 1995 January (Santos-Lle\'{o} et al.\ 1997); 
however, the sampling became markedly worse late in the campaign,
so we restrict our attention here to the data obtained through
early October. The data considered here consist of 39 \IUE\
measurements, with an average spacing of 4.2 days between epochs,
and 25 optical measurements with an average spacing of slightly more
than 6 days.
\item{\bf NGC 4151:} Prior to the NGC 7469 campaign, NGC 4151
represented the state-of-the-art in multiwavelength studies
of AGN variability,
with intensive observations obtained over a 10-day period in
1993. The UV data, with a mean spacing of about 0.05\,day, 
are from Crenshaw et al.\ (1996) and the optical data,
with a mean spacing of about 0.6\,days, are from Kaspi et al.\ (1996).
\item{\bf PKS\,2155--304:} This is an especially well-studied blazar
that was monitored intensively in soft X-rays (Brinkmann et al.\ 1994),
the ultraviolet (Urry et al.\ 1993), and the optical 
(Courvoisier et al.\ 1995) in 1991 November. Analysis of the light curves
by Edelson et al.\ (1995) indicates that the variations in the
soft X-rays precede those in the UV/optical by $\sim$2--3 hours.
\end{enumerate}

For each of these AGNs, we have cross-correlated the data described above
using the interpolation cross-correlation as outlined by
White \& Peterson (1994), using only the parts of the light curves
that overlap in time,  and have estimated
the uncertainties by using the FR/RSS method. The results
for three of the galaxies 
are given in Table 6. In both NGC 5548 monitoring campaigns, 
it was found that the optical continuum (at 5100\,\AA) lagged
behind the UV continuum (at about 1350\,\AA) by $\tau_{cent} \approx
1$\,day.
The uncertainties computed by the FR/RSS method 
indicate that the detected lag is not statistically significant,
as was previously concluded from model-dependent Monte Carlo
simulations. Similar conclusions are reached for both
NGC 3783 and Fairall 9, which were less favorable for detection
of such small lags, and thus have larger associated uncertainties.

The case of NGC 4151 is somewhat more complicated, as shown in
Table 7. The 1275\,\AA/1820\,\AA, 1275\,\AA/2688\,\AA, and
1275\,\AA/5125\,\AA\ cross-correlations have already been
computed by Edelson et al.\ (1996), including maximum
likelihood error estimates $\Delta \tau_{\rm ML}$
on $\tau_{peak}$; these values are
given in column (3) of Table 7. It can be seen that the 
FR/RSS estimates are somewhat smaller for the
1275\,\AA/1820\,\AA\ and  1275\,\AA/2688\,\AA\ cross-correlations,
but similar for 1275\,\AA/5125\,\AA. As concluded by
Edelson et al., the lags for all of these bands relative
to the 1275\,\AA\ continuum are not statistically significant.
However, the 1275\,\AA/6925\,\AA\ cross-correlation, which was {\em not}\
reported by Edelson et al.\ and is reported here for the first time, 
shows that the 6925\,\AA\
continuum lags behind the 1275\,\AA\ continuum by about 2.5 days,
at almost the 90\% confidence level.

The UV/optical CCFs for NGC 4151 have a broad and noisy
structure, and are not very well-defined
(see Fig.\ 5 of Edelson et al.\ 1996). A significant lag between
the UV and optical continuum is thus not obtained, with the possible
exception of the 6925\,\AA\ light curve. The large uncertainties are
mainly a result of the very few optical observations and
the fact that the optical continuum does not
trace the highest UV variability frequencies (Edelson et al.\ 1996).

Given that we have apparently detected a marginally 
significant lag relative to 1275\,\AA\ for only one waveband, it is
worthwhile re-examining all wavelengths in this range, as the existence
of a wavelength-dependent trend in the data may be more convincing,
as indeed it was in the case of the UV data alone in NGC 7469 (see
Fig.\ 6 of Wanders et al.\ 1997). We have done this for both the
UV and optical data by making new measurements from the original
spectra.

All SWP UV spectra of NGC 4151 were rebinned into 20\,\AA-wide bins, and all
LWP UV were rebinned into 50\,\AA-wide
bins. The optical spectra from Wise Observatory were similarly
averaged into 100\,\AA-wide bins.
For each bin, we construct a light curve by measuring
all of the spectra in the time series, and we then cross-correlate these
light curves with the 1275\,\AA\ light curve (for the UV light data)
curves and with the
5125\,\AA\ light curve (for the optical light curves). We then find
the centroid lags as a function of wavelength. The results are plotted
in Fig.~2 for the UV data and in Fig.~3 for the optical data.
For the UV data, Fig.~2 shows that
the lags are consistent with a constant value of zero,
independent of wavelength. For the optical data, 
this is not so obvious. Clearly, there appears to be an
increase in lag with wavelength, at least for wavelengths longer than
about 5100\,\AA. For shorter wavelengths, the light curves are
contaminated with emission-line features, which have a longer lag
than the continuum (Kaspi et al.\ 1996), 
and these wavelengths are not representative
of the true continuum. For the longer wavelengths, we see contamination
by the H$\alpha$ emission line, but other than that, 
there is very little line contamination.

Although the origin of the continuum emission in blazars is almost
certainly different than in other AGNs, we include the results on
the blazar PKS\,2155$-$304 for completeness and as another
illustration of the efficacy of the FR/RSS method. The results
are given in Table 8; note that the uncertainties in this Table
are 90\% confidence limits rather than 68\% confidence limits
for direct comparison with Table 2 of Edelson et al. (1995). Comparison
of the FR/RSS results with those from several other methods 
reveals that the FR/RSS errors are only slightly larger than
those based on  the
$\chi^2$ minimization method, which Edelson et al.\ regard as the most
powerful technique for these particular data. 

In summary, we find that in the cases of NGC 5548, NGC 3783, and
Fairall 9, statistically significant wavelength-dependent continuum
lags are not found, simply because the uncertainties in the
lag determinations are relatively large, greater than about 1 day in
each case. Only in the case of NGC 4151, which is somewhat less luminous
than NGC 7469 and therefore might be expected to show more rapid
variability, is the temporal sampling good enough that one might
hope to see wavelength-dependent continuum lags. And indeed, there is
at least weak
evidence that this effect also does appear in NGC 4151, but only
at wavelengths longer than about 5100\,\AA.
In the case of the blazar PKS\,2155$-$304, the FR/RSS method
confirms the $\sim$2-hr X-ray/UV lag at a high level of confidence.

\section{Discussion}

The results presented here establish the existence of wavelength-dependent
continuum lags in at least one case.  But do the measured lags
indeed represent processes that are occurring in the continuum 
source itself? On account of the large line widths in the 
rich emission-line spectra of AGNs, there are essentially no
truly ``line-free'' continuum windows; all wavelengths are at
least somewhat contaminated by line emission. Is it possible, in fact, that
these time delays are due to emission-line contamination?
Excluding the contamination by the known strong emission lines,
the wavelength-dependence of the lags in NGC 7469
(Fig.\ 6 of Wanders et al.\ 1997 
and Figs.\ 6 and 7 of Collier et al.\ 1998) argue fairly
strongly against contamination by weaker emission features.
Any possible emission-line contaminants would have to show a correlation
of lag with wavelength that would reproduce the approximate
$\tau \propto \lambda^{4/3}$ relationship apparently
detected in NGC 7469 (Collier et al.\ 1998). It remains possible that the 
interband lags we have detected are due to unidentified
contaminants, but we regard this as unlikely and
will therefore assume that the measured lags are indeed
a property of the continuum source. 

If the detected wavelength-dependent continuum lags
arise as a consequence of the temperature structure of
an accretion-disk continuum source,
we need to ask whether or not the detection of
wavelength-dependent continuum lags in NGC 7469 is 
consistent with the absence of detectable lags in NGC 3783, NGC 5548,
and Fairall 9 and with the marginal detection in NGC 4151.
Thus far we have concentrated only on the differences 
in the observations, and have not considered the differing
intrinsic properties of these various galaxies and how
this might influence the detection of such lags. In this
section, we will assume that a simple thermal accretion disk
model can account for the UV/optical continua of non-blazar AGNs,
and we will scale the results for NGC\,7469 to other AGNs
to determine whether or not we should have expected to detect
wavelength-dependent continuum lags in the programs that have
been carried out.

If we assume that the continuum is thermal emission from an
accretion disk, and we further assume that the variability
in different parts of the accretion disk is coupled by
radiative reprocessing, then we find that the continuum
at wavelength $\lambda$ is time-delayed relative to the
primary source of photons such that
\begin{equation}
\tau \propto \left( M \dot{M} \right)^{1/3} \lambda^{4/3},
\end{equation}
where $M$ is the AGN mass and $\dot{M}$ is the accretion rate
(see Collier et al.\ 1998). We further assume that the
UV luminosity is proportional to the accretion rate
(i.e., we assume a constant efficiency for conversion of mass into light),
and that the mass of the central source can be estimated virially
from the broad-line widths $v_{\mbox{\tiny FWHM}}$
and emission-line lags $\tau_{\mbox{\tiny LT}}$, i.e.,
$M \propto v_{\mbox{\tiny FWHM}}^2 \tau_{\mbox{\tiny LT}}$. By making these
simple assumptions and scaling the results relative to 
NGC 7469, we can compute the expected UV-to-optical lag
$\Delta \tau$ under the assumption
that $\dot{M}$ scales with the UV luminosity and that
$M$ scales like the square of the Ly$\alpha$ line width
times its time delay. We show the results of this na\"{\i}ve calculation
in Table 9. Column (1) gives the galaxy name, and columns
(2) and (3) give $M$ and $\dot{M}$, respectively, scaled to NGC 7469
assuming the relationships given above. These are used to compute
a time delay $\Delta \tau$ between the
UV and optical wavelengths given in columns (4) and (5), respectively.
This predicted time delay, scaled to the
1.8-day delay between the 1315\,\AA\ and 6962\,\AA\ variations in
NGC 7469, is given in column (7). The expected lag given in
column (7) should be compared with the 90\% confidence interval
(two-sided) on $\tau_{cent}$ computed by the FR/RSS method,
which is given in column (8). In the case of NGC 4151, the 90\% confidence
interval for the 1275\,\AA/6925\,\AA\ cross-correlation is
surprisingly broad compared to the $\pm\Delta\tau_{68}$ width
of only 0.623\,days (for the centroid). This is because of the
highly non-normal CCPD, as shown in Fig.~4. 
In this case, the CCPD has a extended 
low-level tail, which is a characteristic of CCPDs based on a small
number of data points (there are only 9 points in the
6925\,\AA\ light curve).

In each of the cases studied here, we find that the non-detection
(or very marginal detection in the case of NGC 4151)
of wavelength-dependent continuum lags is consistent with the
predictions of the simple scaling model. The lags predicted by
the scaling adopted are simply too small to have been detected by
the experiments that have been carried out. Both the predicted
lags based on scaling relative to  NGC 7469 and zero lag
lie safely within the 90\% confidence interval in each case.

Given this na\"{\i}ve model and the considerable
uncertainties in the determination of both $M$ and $\dot{M}$, 
we do not believe that previous non-detection of 
wavelength-dependent continuum lags poses any new 
theoretical challenge to our understanding of accretion disks
in AGNs.

\section{Conclusions}

In this paper, we have introduced a model-independent Monte-Carlo
method for estimating conservatively 
uncertainties in cross-correlation lags, and have
tested this method for a variety of conditions. 
We have applied this method to the UV/optical light curves of
several well-studied non-blazar AGNs and the blazar PKS\,2155$-$304.
The reality of wavelength-dependent
continuum lags found in  NGC 7469 (Wanders et al.\ 1997;
Collier et al.\ 1998) is supported by this analysis
at at level of confidence higher than 95\%.
We confirm at about a 95\% confidence level the detection of a
$\sim$2-hr X-ray/UV lag in PKS\,2155$-$304.
However, we find no evidence for statistically significant 
UV/optical continuum lags in the non-blazar AGNs
NGC 5548, NGC 3783, and Fairall 9, and we find only marginal
evidence for the effect in NGC 4151. By scaling other AGNs relative to
NGC 7469, we find however that the absence of significant
wavelength-dependent continuum lags in other AGNs does not appear to be
inconsistent with a thermal accretion-disk model --- most
AGNs have not been monitored at high enough temporal frequency
for UV/optical continuum lags to be detectable. Failure to detect 
such lags in previous experiments does not pose a serious
threat to accretion-disk models, as long as the variability
signal is propagated at the speed of light.

We are grateful for support of this study by
NASA (through ADP grant NAG5-3497 and LTSA grant NAG5-3233) and by
the National Science Foundation (through grant AST94-20080).
The authors thank P.T.\ O'Brien and an anonymous referee
for a number of suggestions.
\clearpage



\clearpage

\begin{figure}
\caption{Cross-correlation peak distributions for 
an FR/RSS simulation. The specific example shown is
Test 2 in Table 2, i.e., the response of a thin spherical
shell of radius $R = 20$\,light days. Note that the
distribution of CCFs peak values ($\tau_{peak}$, upper
panel) is broader than the distribution for the 
CCF centroid ($\tau_{cent}$, lower panel). We find this
to be generally the case.}
\end{figure}

\begin{figure}
\caption{The UV time lags as a function of wavelength for NGC\,4151.
The upper panel shows the average SWP (shorter wavelengths) spectrum
and the average LWP (longer-wavelength spectrum). The second panel
shows the root-mean-square (RMS) spectrum, in which the most
variable parts of the spectrum stand out. The third panel
shows the maximum correlation coefficient of the CCF $r_{max}$ for
each wavelength bin, and the bottom panel shows the CCF
centroid $\tau_{cent}$ of the CCF that results from cross-correlating
the measurements at each particular wavelength with respect to the
1300\,\AA\ light curve. In the bottom panel, the highest peaks
are due to the relatively slow response of Ly$\alpha\,\lambda1216$
and C\,{\sc iv}\,$\lambda1549$. Other than these features, there
is no apparent wavelength-dependent trend as there was in
NGC 7469 (Wanders et al.\ 1997).}
\end{figure}

\begin{figure}
\caption{The optical time lags as a function of wavelength for NGC\,4151.
The upper panel shows the average blue and red spectra obtained at
Wise Observatory. The second panel shows the RMS spectra
computed from these same data. The light curves for each wavelength
bin are cross-correlated with the 5125\,\AA\ light curve;
the third panel shows the peak value of the cross-correlation
coefficient $r_{max}$ as a function of wavelength, and
the centroid $\tau_{cent}$ for each wavelength is shown in 
the bottom panel. The highest peak in the bottom panel is
due to the slow response of the H$\alpha$ emission line. There
is also a hint of the He\,{\sc ii}\,$\lambda4686$ response.
The bottom panel shows that $\tau_{cent}$ appears to increase with
wavelength longward of about 5100\,\AA. This appears to be
similar to what was seen in NGC 7469.}
\end{figure}

\begin{figure}
\caption{Cross-correlation peak distributions for 
the 1275\,\AA/6925\,\AA\ continuum bands in NGC 4151.
The CCPD shows a strong peak at at time delay of about
2.5 days, but slightly more than 10\% of the realizations
yield values that form an extended tail towards lags smaller than
about 2 days. This occurs on
account of the very small number (9) of data points
in the 6925\,\AA\ light curve; with so few points, 
it is generally not possible to determine a lag at
a high level of confidence.}
\end{figure}


\clearpage

\begin{deluxetable}{lllll} 
\tablewidth{0pt}
\tablecaption{A Typical Test}
\tablehead{
\colhead{Parameter}                &
\colhead{MC}                       &
\colhead{FR/RSS}                   &
\colhead{RSS}                      &
\colhead{FR}                       \nl
\colhead{(1)} &
\colhead{(2)} &
\colhead{(3)} &
\colhead{(4)} &
\colhead{(5)} 
}
\startdata
$\tau_{cent}$ (days)&
$19.1 ^{+1.9}_{-1.9}$ &
$21.0 ^{+2.5}_{-2.7}$ &
$21.0 ^{+1.5}_{-2.5}$ &
$21.0 ^{+1.5}_{-1.3}$ \nl
$\tau_{peak}$ (days) &
$20^{+2} _{-2}$ &
$20^{+2} _{-2}$ &
$20^{+1} _{-2}$ &
$20^{+1} _{-2}$ \nl
\enddata
\end{deluxetable}

\begin{deluxetable}{ccccc}
\tablewidth{0pt}
\tablecaption{Repeated Tests with Constant Parameters ---
Thin-Shell Model}
\tablehead{
\colhead{Test} &
\multicolumn{2}{c}{$\tau_{cent}$ (days)}  &
\multicolumn{2}{c}{$\tau_{peak}$ (days)}  \nl
\colhead{No.}			   &
\colhead{MC}                       &
\colhead{FR/RSS}                   &
\colhead{MC}                       &
\colhead{FR/RSS}                   \nl
\colhead{(1)} &
\colhead{(2)} &
\colhead{(3)} &
\colhead{(4)} &
\colhead{(5)} 
}
\startdata
1 & $19.1 ^{+1.9}_{-1.9}$ & $21.0 ^{+2.5}_{-2.7}$ &
$20^{+2} _{-2}$ & $20^{+2} _{-2}$ \nl

2 & $17.1 ^{+2.6}_{-2.4}$ & $16.2 ^{+4.5}_{-2.6}$ &
$20^{+4} _{-4}$ & \hspace*{4pt}$24^{+1} _{-12}$ \nl

3 & $19.5 ^{+2.7}_{-2.6}$ & $23.9 ^{+3.3}_{-7.0}$ &
$20^{+4} _{-4}$ & \hspace*{4pt}$24^{+2} _{-12}$ \nl

4 & $20.0 ^{+1.5}_{-1.6}$ & $18.4 ^{+2.0}_{-2.3}$ &
$20^{+2} _{-2}$ & $18^{+3} _{-2}$ \nl

5 & $18.0 ^{+2.5}_{-2.7}$ & $18.7 ^{+3.7}_{-5.6}$ &
$21^{+5} _{-6}$ & $21^{+6} _{-5}$ \nl

6 & $20.2 ^{+2.8}_{-2.3}$ & $24.0 ^{+3.9}_{-3.9}$ &
$20^{+3} _{-3}$ & $22^{+8} _{-2}$ \nl

7 & $18.9 ^{+2.5}_{-2.6}$ & $16.1 ^{+3.7}_{-3.1}$ &
$19^{+2} _{-3}$ & $16^{+7} _{-3}$ \nl

8 & $19.1 ^{+1.9}_{-1.9}$ & $19.6 ^{+1.9}_{-2.9}$ &
$19^{+3} _{-2}$ & $21^{+2} _{-3}$ \nl

9 & $22.3 ^{+3.9}_{-4.2}$ & $17.6 ^{+8.9}_{-4.1}$ &
$21^{+3} _{-4}$ & $19^{+4} _{-5}$ \nl

10 & \hspace*{4pt}$18.5 ^{+10.5}_{-8.7}$ & $20.0 ^{+6.4}_{-8.3}$ &
$20^{+5} _{-5}$ & $21^{+6} _{-3}$ \nl

\enddata
\end{deluxetable}

\begin{deluxetable}{ccccc}
\tablewidth{0pt}
\tablecaption{Repeated Tests with Constant Parameters---
Thick-Shell Model}
\tablehead{
\colhead{Test} &
\multicolumn{2}{c}{$\tau_{cent}$ (days)}  &
\multicolumn{2}{c}{$\tau_{peak}$ (days)}  \nl
\colhead{No.}			   &
\colhead{MC}                       &
\colhead{FR/RSS}                   &
\colhead{MC}                       &
\colhead{FR/RSS}                   \nl
\colhead{(1)} &
\colhead{(2)} &
\colhead{(3)} &
\colhead{(4)} &
\colhead{(5)} 
}
\startdata
1 & $11.56 ^{+0.94}_{-0.81}$ & $10.77 ^{+1.27}_{-1.24}$ &
$12.0^{+1.0} _{-1.0}$ & $11.0^{+1.0} _{-1.0}$ \nl

2 & $10.31 ^{+1.41}_{-1.67}$ & \mone$9.55 ^{+1.25}_{-1.24}$ &
$11.0^{+2.0}_{-2.5}$ & $10.5^{+1.0} _{-1.5}$ \nl

3 & $13.85 ^{+1.47}_{-1.66}$ & $11.76 ^{+2.02}_{-1.19}$ &
$12.0^{+2.0} _{-1.5}$ & $14.0^{+0.5} _{-3.0}$ \nl

4 & $10.80 ^{+0.90}_{-1.00}$ & $11.00 ^{+0.81}_{-1.03}$ &
$11.0^{+1.0} _{-1.5}$ & $12.0^{+0.5} _{-3.0}$ \nl

5 & $12.19 ^{+1.51}_{-1.24}$ & $12.38 ^{+1.07}_{-3.03}$ &
$12.0^{+1.5} _{-2.0}$ & $12.5^{+1.0} _{-3.5}$ \nl

6 & \mone$9.96 ^{+1.75}_{-1.90}$ & $\mone9.87 ^{+2.63}_{-2.92}$ &
$11.0^{+2.5} _{-2.5}$ & $11.5^{+3.0} _{-4.0}$ \nl

7 & \mone$8.98 ^{+1.13}_{-1.17}$ & $\mone9.75 ^{+1.95}_{-0.83}$ &
$10.0^{+1.5} _{-2.0}$ & \mone$9.0^{+4.0} _{-1.5}$ \nl

8 & \mone$9.56 ^{+1.52}_{-1.43}$ & \mone$8.58 ^{+2.58}_{-2.52}$ &
$11.0^{+2.0} _{-2.0}$ & \mone$9.0^{+2.0} _{-1.5}$ \nl

9 & \mone$6.42 ^{+1.80}_{-1.74}$ & \mone$9.05 ^{+1.48}_{-2.35}$ &
\mone$8.5^{+3.0} _{-3.5}$ & \mone$8.5^{+6.0} _{-1.0}$ \nl

10 & $11.85 ^{+0.91}_{-0.91}$ & $12.25 ^{+1.67}_{-0.84}$ &
$11.5^{+1.5} _{-1.0}$ & $11.0^{+2.5} _{-0.5}$ \nl
\enddata
\end{deluxetable}

\begin{deluxetable}{lcllllc} 
\tablewidth{0pt}
\tablecaption{Thin-Shell Models in the Poor-Sampling Limit}
\tablehead{
\colhead{PDS}                &
\colhead{$R$}    &
\colhead{ }	&
\multicolumn{2}{c}{$\tau_{cent}$ (days)}   &
\multicolumn{2}{c}{$\tau_{peak}$ (days)}   \nl
\colhead{index} &
\colhead{(lt days)} &
\colhead{$N$}  &
\colhead{MC} &
\colhead{FR/RSS}   &
\colhead{MC} &
\colhead{FR/RSS}   \nl
\colhead{(1)} &
\colhead{(2)} &
\colhead{(3)} &
\colhead{(4)} &
\colhead{(5)}&
\colhead{(6)}&
\colhead{(7)}
}
\startdata
$-2.5$ & $20$ & $40$ &
	$18.5 ^{+3.0} _{-4.6}$ &
	$14.8 ^{+6.8} _{-0.4}$ &
	$19^{+4}_{-4}$ &
	$19^{+5}_{-4}$ \nl
$-2.5$ & $20$ & $20$ &
	$19.8 ^{+3.5} _{-5.1}$ &
	$17.8 ^{+5.3} _{-6.4}$ &
	$19 ^{+6} _{-5}$ &
	$19 ^{+6} _{-8}$ \nl
$-2.5$ & $20$ & $10$ &
	$19.9 ^{+5.6} _{-4.5}$ &
	$18.6 ^{+7.5} _{-6.5}$ &
	$19 ^{+6} _{-4}$ &
	$20 ^{+6} _{-9}$ \nl
$-2.5$ & $2$ & $40$ &
	\mone$2.0 ^{+0.9} _{-1.0}$ &
	\mone$1.9 ^{+2.0} _{-1.9}$ &
	\mone$2 ^{+1} _{-1}$ &
	\mone$2 ^{+1} _{-2}$ \nl
$-2.5$ & $2$ & $20$ &
	\mone$1.9 ^{+1.0} _{-1.0}$ &
	\mone$1.5 ^{+1.5} _{-2.5}$ &
	\mone$2 ^{+1} _{-1}$ &
	\mone$2 ^{+1} _{-3}$ \nl
$-2.5$ & $2$ & $10$ &
	\mone$0.5 ^{+5.2} _{-4.2}$ &
	\mone$5.7 ^{+10.4} _{-15.7}$ &
	\mone$1 ^{+4} _{-5}$ &
	\hspace*{9pt}$4 ^{+7} _{-20}$ \nl
$-1.8$ & $20$ & $40$ &
	$18.3 ^{+3.4} _{-3.5}$ &
	$22.5 ^{+4.1} _{-6.3}$ &
	$19 ^{+7} _{-7}$ &
	$22 ^{+8} _{-8}$ \nl
$-1.8$ & $20$ & $20$ &
	$19.4 ^{+5.5} _{-4.6}$ &
	$18.0 ^{+5.8} _{-2.9}$ &
	$21 ^{+6} _{-9}$ &
	$\hspace*{4pt}14 ^{+11} _{-7}$ \nl
$-1.8$ & $20$ & $10$ &
	$18.5 ^{+8.5} _{-7.4}$ &
	$17.6 ^{+11.7} _{-8.3}$ &
	$19 ^{+8} _{-7}$ &
	$\hspace*{4pt}20 ^{+17} _{-9}$ \nl
\enddata
\end{deluxetable}

\begin{deluxetable}{lllclc} 
\tablewidth{0pt}
\tablecaption{NGC 7469 Cross-Correlation Results}
\tablehead{
\colhead{ } 			&
\colhead{ }			&
\multicolumn{2}{c}{Centroid}        &
\multicolumn{2}{c}{Peak}    \nl
\colhead{First}                   &
\colhead{Second}                  &
\colhead{$\tau_{cent}$}           &
\colhead{ }			  &
\colhead{$\tau_{peak}$}           &
\colhead{ }			  \nl
\colhead{Series}                  &
\colhead{Series}                  &
\colhead{(days)}                  &
\colhead{$P(\leq0\,{\rm days})$}  &
\colhead{(days)}                  &
\colhead{$P(\leq0\,{\rm days})$}  \nl
\colhead{(1)}                &
\colhead{(2)}                &
\colhead{(3)}                &
\colhead{(4)}                &
\colhead{(5)}                &
\colhead{(6)}                 
}
\startdata
1315\,\AA\ &1825\,\AA\ 	& \mm0.22$^{+0.12}_{-0.13}$ & 0.024
			& \mm0.07$^{+0.19}_{-0.12}$ & 0.166 \nl
1315\,\AA\ &4845\,\AA\ 	& \mm1.25$^{+0.48}_{-0.35}$ & 0.004
			& \mm1.30$^{+0.45}_{-0.45}$ & 0.006 \nl
1315\,\AA\ &6962\,\AA\ 	& \mm1.84$^{+0.93}_{-0.94}$ & 0.030
			& \mm1.40$^{+1.50}_{-0.30}$ & 0.046 
\enddata
\end{deluxetable}

\begin{deluxetable}{llclc} 
\tablewidth{0pt}
\tablecaption{UV/Optical Continuum Lags for Well-Studied AGNs}
\tablehead{
  & \multicolumn{2}{c}{Centroid} 
 & \multicolumn{2}{c}{Peak} \nl
\colhead{ }                   &
\colhead{$\tau_{cent}$}       &    
\colhead{ }                   &
\colhead{$\tau_{peak}$}       &
\colhead{ }                   \nl
\colhead{Galaxy} & 
\colhead{(days)}                  &
\colhead{$P(\leq0\,{\rm days})$}  &
\colhead{(days)}                  &
\colhead{$P(\leq0\,{\rm days})$}  \nl
\colhead{(1)}      	     & 
\colhead{(2)}                &
\colhead{(3)}                &
\colhead{(4)}                &
\colhead{(5)}                
}
\startdata
NGC 5548 {\it IUE}\ 1989\tablenotemark{a}
                   & \mm1.3$^{+1.5}_{-1.7}$  & 0.351 
                   & \mm2.5$^{+1.5}_{-3.0}$  & 0.205 \nl
NGC 5548 {\it HST/IUE}\ 1993\tablenotemark{b}
		   & \mm1.2$^{+0.6}_{-1.2}$  & 0.191
                   & \mm0.7$^{+1.3}_{-1.2}$  & 0.341 \nl
NGC 3783\tablenotemark{c}
	           & \mm$0.2^{+2.2}_{-2.0}$  & 0.478
                   & \mm0.9$^{+0.7}_{-1.8}$  & 0.354 \nl
Fairall 9\tablenotemark{d}
	           & $-2.8^{+6.5}_{-2.5}$    & 0.622
                   & $-0.5^{+7.5}_{-5.5}$    & 0.584 \nl
\enddata
\tablenotetext{a}{\raggedright UV: 1337\,\AA\ flux (GEX version)
from Clavel et al.\ (1991), Table 2. Optical: 5100\,\AA\ flux
from Peterson et al.\ (1992), Table 12.}
\tablenotetext{b}{\raggedright From Korista et al.\ (1995).
UV: 1350\,\AA\ flux from Table 12, plus selected {\it IUE}\ SWP
fluxes before JD 2449079.0 from Table 16. Optical: 5100\,\AA\
fluxes from Table 21.} 
\tablenotetext{c}{\raggedright UV: 1460\,\AA\ flux (GEX version)
from Reichert et al.\ (1994), Table 5. Optical: 5150\,\AA\ flux
from Stirpe et al.\ (1994), Table 7.}
\tablenotetext{d}{\raggedright UV: 1390\,\AA\ flux 
from Rodr\'{\i}guez-Pascual et al.\ (1997), Table 2.
Optical: 5340\,\AA\ flux from Santos-Lle\'{o} et al.\ (1997),
Table 2.}
\end{deluxetable}

\begin{deluxetable}{llcccccc} 
\tablewidth{0pt}
\tablecaption{NGC 4151 Cross-Correlation Results}
\tablehead{
 & & \multicolumn{3}{c}{Peak} && \multicolumn{2}{c}{Centroid} \nl
\cline{3-5}\cline{7-8}
\colhead{First}                   &
\colhead{Second}                 &
\colhead{$\tau_{peak} \pm \Delta \tau_{\rm ML}$\tablenotemark{a}}  &
\colhead{$\tau_{peak}$}           &
\colhead{ }			  &
\colhead{ }			  &
\colhead{$\tau_{cent}$}           &
\colhead{ }			  \nl
\colhead{Series}                  &
\colhead{Series}                  &
\colhead{(days)}                  &
\colhead{(days)}                  &
\colhead{$P(\leq0\,{\rm days})$}  &
\colhead{ }			  &
\colhead{(days)}                  &
\colhead{$P(\leq0\,{\rm days})$}  \nl
\colhead{(1)}                &
\colhead{(2)}                &
\colhead{(3)}                &
\colhead{(4)}                &
\colhead{(5)}                &
\colhead{ }			  &
\colhead{(6)}                &
\colhead{(7)}             
}
\startdata
1275\,\AA\ &1820\,\AA\ 	& $-$0.01$^{+0.06}_{-0.09}$ 
			& $-$0.025$^{+0.005}_{-0.040}$ & 0.930 &
			& $-$0.049$^{+0.028}_{-0.023}$ & 0.991 \nl
1275\,\AA\ &2688\,\AA\ 	& \mm0.00$^{+0.20}_{-0.07}$ 
			& \mm0.000$^{+0.095}_{-0.005}$ & 0.470&
			& \mm0.086$^{+0.039}_{-0.048}$ & 0.042 \nl
1275\,\AA\ &5125\,\AA\ & $-$0.05$^{+0.58}_{-0.73}$ 
			& \mm0.430$^{+0.220}_{-1.269}$ & 0.449 &
			& \mm0.296$^{+0.307}_{-0.938}$ & 0.423 \nl
1275\,\AA\ &6925\,\AA\ & $\ldots$
			& \mm2.485$^{+0.280}_{-0.450}$ & 0.129&
			& \mm2.462$^{+0.315}_{-0.308}$ & 0.102 \nl
\enddata
\tablenotetext{a}{From Edelson et al.\ (1996) Table 4.}
\end{deluxetable}

\begin{deluxetable}{llcccc} 
\tablewidth{0pt}
\tablecaption{PKS 2155$-$304 Cross-Correlation Results}
\tablehead{
 & & \multicolumn{2}{c}{Centroid} & \multicolumn{2}{c}{Peak} \nl
\colhead{First}                   &
\colhead{Second}                 &
\colhead{$\tau_{cent}$\tablenotemark{a}}&
\colhead{ }                      &
\colhead{$\tau_{peak}$\tablenotemark{a}} &
\colhead{ }                      \nl
\colhead{Series}                  &
\colhead{Series}                  &
\colhead{(hrs)}                  &
\colhead{$P(\leq0\,{\rm hrs})$}   &
\colhead{(hrs)}                   &
\colhead{$P(\leq0\,{\rm hrs})$}   \nl
\colhead{(1)}                &
\colhead{(2)}                &
\colhead{(3)}                &
\colhead{(4)}                &
\colhead{(5)}                &
\colhead{(6)}                 
}
\startdata
24\,\AA\ & 1400\,\AA\   &\mm$1.49^{+1.12}_{-1.37}$ & 0.052
			&\mm$2.2^{+1.0}_{-1.6}$ & 0.034 \nl
1400\,\AA\ & 2800\,\AA\ &\mm$0.05^{+1.65}_{-1.15}$ & 0.520
			&\mm$0.1^{+1.2}_{-1.2}$ & 0.621 \nl
2800\,\AA\ & 5000\,\AA\	&$-0.70^{+2.72}_{-2.64}$ & 0.729
			&$-0.6^{+3.0}_{-3.4}$ & 0.676 \nl
\enddata
\tablenotetext{a}{Confidence levels are 90\% for direct comparison
with Edelson et al.\ (1995) Table 2.}
\end{deluxetable}

\begin{deluxetable}{lllllcl} 
\tablewidth{0pt}
\tablecaption{Comparison with Theoretical Scaling}
\tablehead{
\colhead{ }                      &
\colhead{ }        &
\colhead{ }  &
\colhead{$\lambda_{1}$}		      &
\colhead{$\lambda_{2}$}		      &
\colhead{$\Delta \tau$}   &
\colhead{90\% Confidence}  \nl
\colhead{Galaxy}                      &
\colhead{$M$\tablenotemark{a}}        &
\colhead{$\dot{M}$\tablenotemark{a}}  &
\colhead{(\AA)}		      &
\colhead{(\AA)}		      &
\colhead{(days)}   &
\colhead{Interval (days)}  \nl
\colhead{(1)}                &
\colhead{(2)}                &
\colhead{(3)}                &
\colhead{(4)}                &
\colhead{(5)}		     &
\colhead{(7)}		     &
\colhead{(8)}
}
\startdata
NGC 4151 & 0.53 & 0.44 
         &    1275 & 6925 & 1.1  & $[-1.90,+3.00]$ \nl
         &&&  1275 & 5125 & 0.7 & $[-1.25, +1.40]$ \nl
NGC 3783 & 2.92 & 0.38 
	 &    1460 & 5150 & 1.1  & $[-4.69, +4.62]$\nl
NGC 5548 & 2.00 & 1.04
	 &    1350 & 5150 & 1.4  & $[-0.54, +2.49]$ \nl
Fairall 9& 4.74 & 8.42 
	 &    1390 & 5340 & 4.0  & $[-9.63, +10.28]$ \nl
\enddata
\tablenotetext{a}{Relative to NGC 7469.}
\end{deluxetable}



\begin{references}
\reference{} 
Alexander, T. 1997,  in
Astronomical Time Series, ed.\ D.\ Maoz, A.\ Sternberg, \& E.M.\ Leibowitz,
(Dordrecht: Kluwer Academic Publishers), p.\ 163

\reference{}
Brinkmann, W., et al. 1994, A\&A, 288, 433

\reference{} 
Collier, S., et al. 1998, ApJ, 498, in press

\reference{}
Courvoisier, T.J.-L., et al. 1995, ApJ, 438, 108

\reference{} 
Clavel, J., et al. 1991, ApJ, 366, 64

\reference{} 
Crenshaw, D.M., et al. 1996, ApJ, 470, 322

\reference{} 
Edelson, R.A., \& Krolik, J.H. 1988, ApJ, 333, 646

\reference{} 
Edelson, R.A., et al. 1995, ApJ, 438, 120

\reference{} 
Edelson, R.A., et al. 1996, ApJ, 470, 364

\reference{}
Gaskell, C.M., \& Peterson, B.M. 1987, ApJS, 65, 1

\reference{} 
Gaskell, C.M., \& Sparke, L.S. 1986, ApJ, 305, 175

\reference{} 
Kaspi, S., et al. 1996, ApJ, 470, 336

\reference{} 
Koratkar, A.P., \& Gaskell, C.M. 1991, ApJS, 75, 719

\reference{} 
Korista, K.T., et al. 1995, ApJS, 97, 285

\reference{} 
Maoz, D., \& Netzer, H. 1989, MNRAS, 36, 21

\reference{}
Mineshige, S., \& Shields, G.A. 1990, ApJ, 351, 47

\reference{} 
Netzer, H., \& Peterson, B.M. 1997, in
Astronomical Time Series, ed.\ D.\ Maoz, A.\ Sternberg, \& E.M.\ Leibowitz,
(Dordrecht: Kluwer Academic Publishers), p.\ 85

\reference{} 
Penston, M.V. 1991, in Variability of Active Galactic Nuclei,
ed.\ H.R. Miller \& P.J. Wiita (Cambridge: Cambridge University Press),
p.\ 343

\reference{}
Peterson, B.M. 1993, PASP, 105, 247

\reference{} 
Peterson, B.M., et al. 1991, ApJ, 368, 119

\reference{} 
Peterson, B.M., et al. 1992, ApJ, 392, 469

\reference{}
Peterson, B.M., Wanders, I., Bertram, R., Hunley, J.F., Pogge, R.W.,
\& Wagner, R.W. 1998, ApJ, in press

\reference{}
Press, W.H., Teuloksky, S.A., Flannery, B.P., \& Vettering, W.T. 1992,
{\it Numerical Recipes: The Art of Scientific Computing}, 2nd edition,
(Cambridge: Cambridge University Press), p.\ 686.

\reference{} 
Reichert, G.A., et al. 1994, ApJ, 425, 582

\reference{}
Robinson, A., and P\'{e}rez, E. 1990, MNRAS, 244, 138

\reference{} 
Rodr\'{\i}guez-Pascual, P.M., et al. 1997, ApJS, 110, 9

\reference{} 
Santos-Lle\'{o}, M., et al. 1997, ApJS, 112, 271

\reference{}
Siemiginowska, A., Czerny, B., \& Kostyunin, V. 1996, ApJ, 458, 491

\reference{} 
Stirpe, G.M., et al. 1994, ApJ, 425, 609

\reference{}
Urry, C.M., et al. 1993, ApJ, 411, 614

\reference{} 
Wanders, I., et al., 1995, ApJ, 453, L87

\reference{} 
Wanders, I., et al., 1997, ApJS, 113, 69

\reference{} 
Warwick, R.S., et al. 1996, ApJ, 470, 349

\reference{} 
White, R.J., \& Peterson, B.M. 1994, PASP, 106, 879 

\end{references}
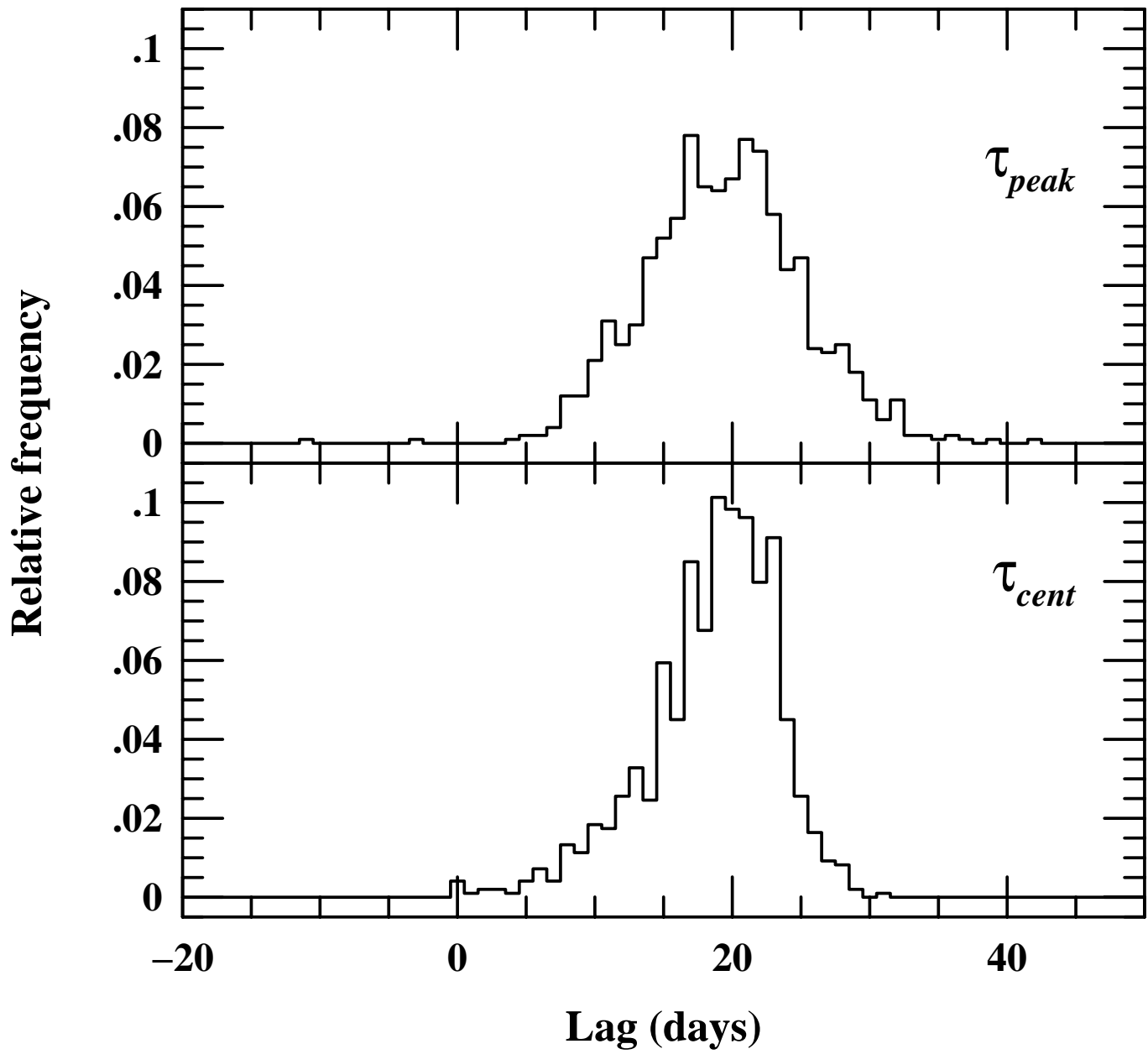
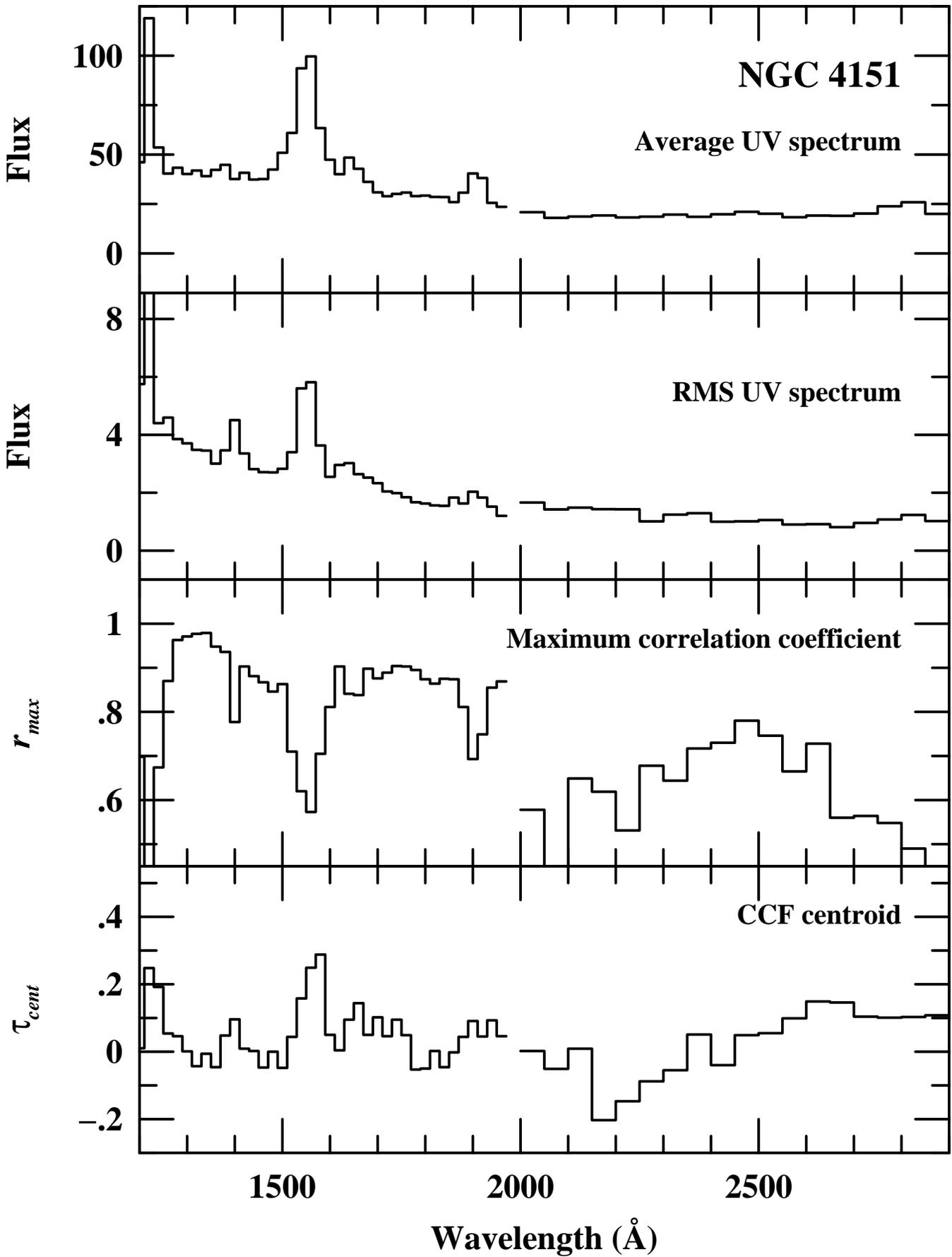
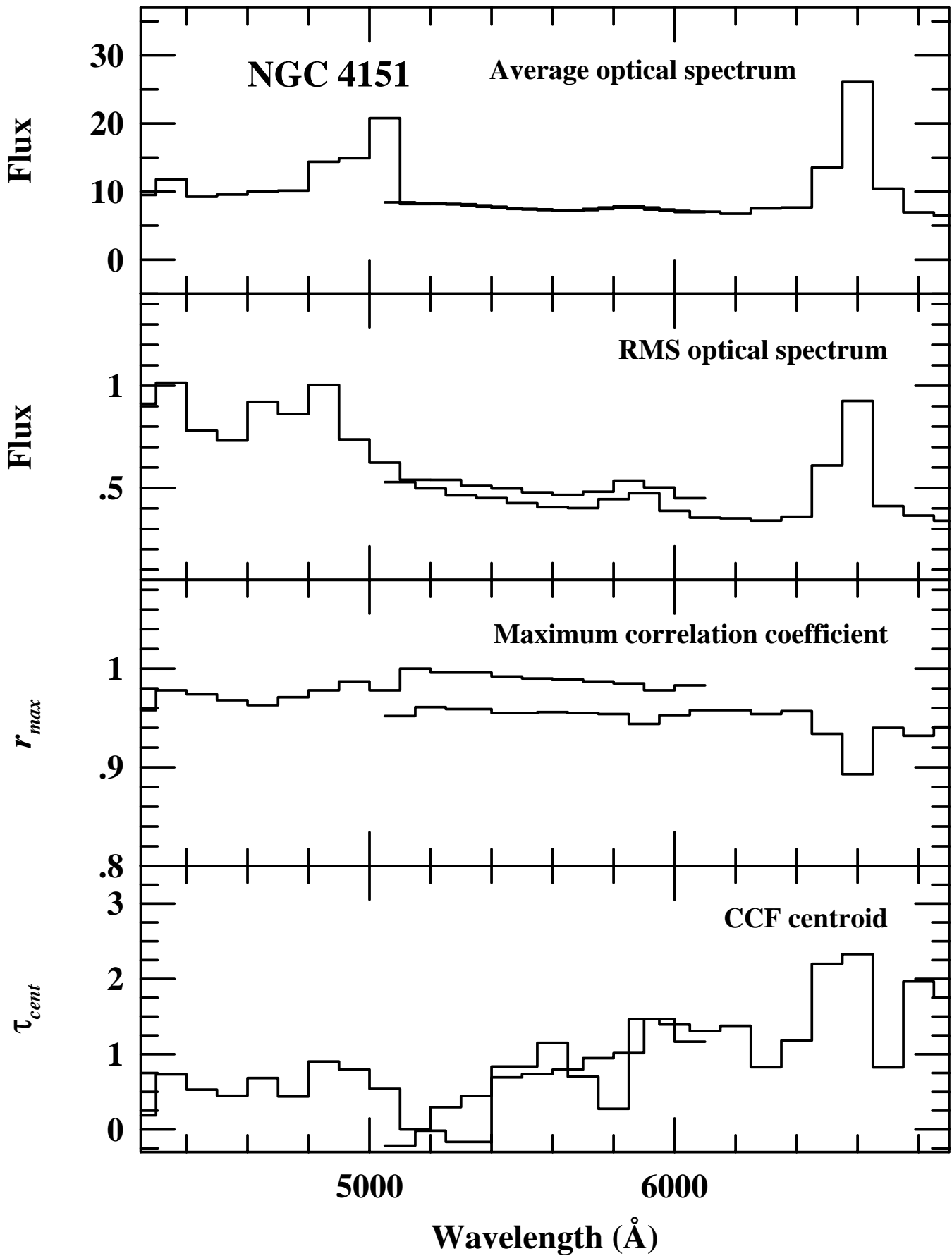
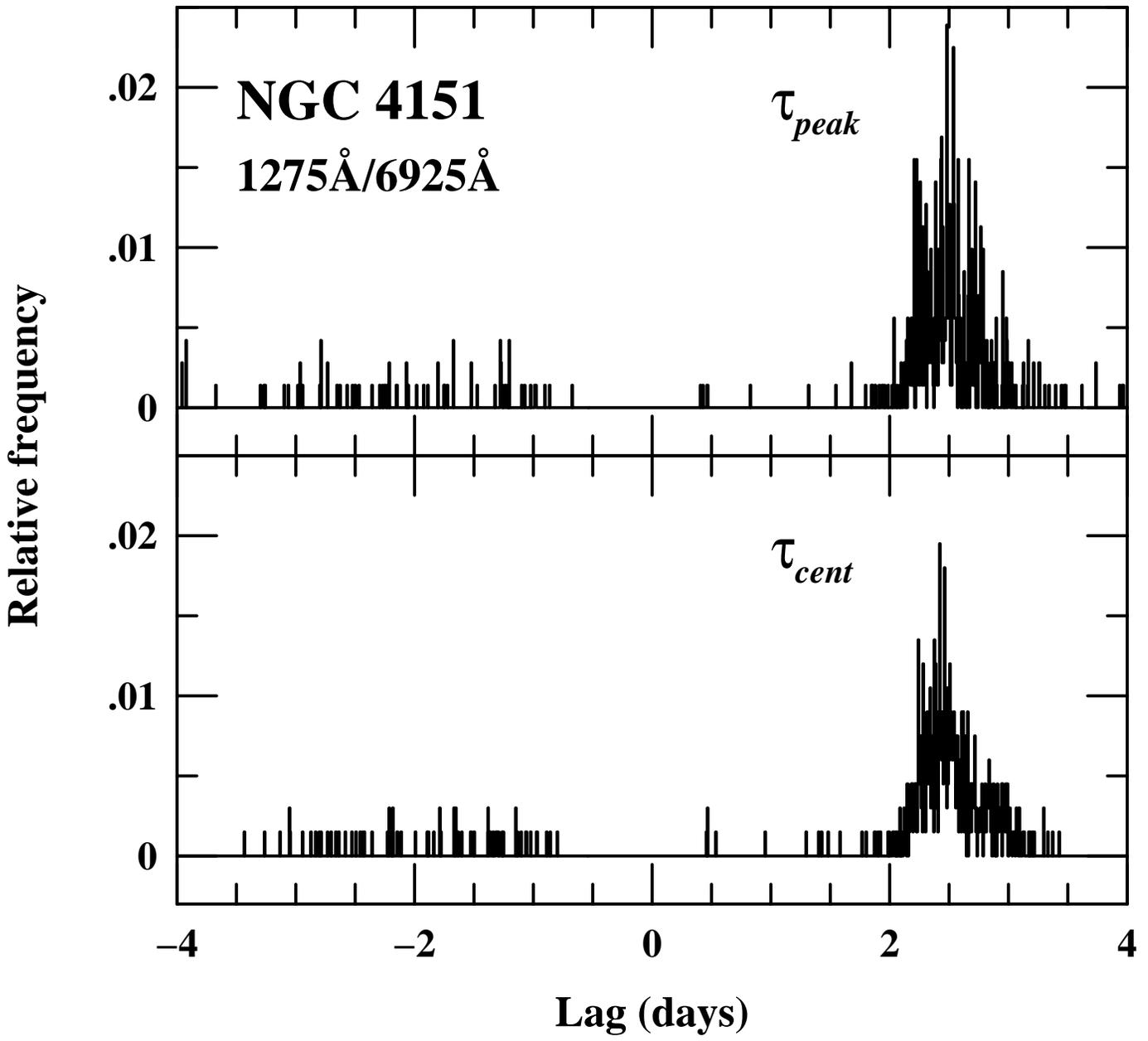
\end{document}